# Configurational Approaches to Urban Form: Empirical Test on the City of Nice (France)

Giovanni Fusco[a], Michele Tirico[b]

[a,b] UMR ESPACE, CNRS / University of Nice Sophia Antipolis, Nice, France.
( [a] giovanni.fusco@unice.fr, [b] michele.tirico@etu.unice.fr ).

Key-words: Configurational Analysis, Urban Form, Space Syntax, Multiple Centrality Assessment, Mark Point Parameter Analysis, Nice

## Introduction

Configurational analysis can be seen as the adaptation to urban space of complex network approaches already developed in social network analysis (Freeman 1979). Its central idea is that elements of urban form should be analysed with respect to the relationships that they establish with all other form elements within a given scale of analysis. In twenty years of research on urban form using configurational analysis, a certain variety of approaches and of more specific techniques has been developed. The main goal of this paper is to give a systematic overview of the different techniques and approaches and to test selected techniques in their ability to recognise different urban forms within the city of Nice, France.

## Methodology

Configurational analysis is a street network approach to the analysis of urban form. Compared to the classical approaches of urban morphology (Caniggia and Maffei 1979, Conzen 1960, Castex et al. 1980), it privileges the analysis of the network of urban streets over the analysis of the built-up areas or of the parcel structure of land. Thanks to the assumptions of the theory of natural movement (Hiller and Hanson 1984), it bases the relations between form elements on the linkages created by the potential movement of pedestrians (usually shortest paths) within urban space. By doing so, configurational analysis goes beyond the geometrical relations linking a form element to its immediate neighbours and determines its configurational properties by taking into account the relations established with all other elements within a given scale of analysis. This allows a multiscale network-based analysis of urban fabric and of overall city layout (Levy 2005).

At first view, scientific literature identifies two broad families of approaches to configurational analysis (Porta et al. 2006a, 2006b). The first is based on metric distances on physical networks and uses a primal graph of the street network, where street intersections are nodes and street segments are edges. Multiple Centrality Assessment (MCA, Porta et al. 2006a) is the most widely used technique of the primal approach. The second approach is based on topological distances on a dual graph of the street network, where linear features (street segments, visual lines, etc.) are nodes and their intersections are edges. Axial analysis of Space Syntax (Hillier and Hanson 1984, Hillier 1996) is a typical example of the dual approach.

According to Porta et al. (2006a) and Ratti (2004) the primal approach should be preferred for configurational analysis. Once problems of edge effects are solved, the primal approach identifies central features within the street networks by means of calculi which are independent from form elements identification, produces indicators which better describe different aspects of centrality and integrates metric distance in the calculus, respecting one of the main properties of physical urban space. For Porta et al. (2006b), the dual approach has great potential in the analysis of street networks but should not be used for configurational calculus, as purported by Hillier (1996).

It seems to us that the differentiation between a primal and a dual model of the street network is a main difference between configurational techniques. But it is not the only one. Two other main methodological aspects should be considered. The first one concerns the entitization of the network, i.e. the way we identify the physical elements of the street network. There are at least four different options: topological elements (defined by connections which are consistent through deformation and magnification/ reduction of the urban space), angular elements (defined by directions which are consistent through magnification/reduction but won't resist deformation), dimensional elements (segments of a given length, which won't resist neither deformation nor magnification/reduction), socially defined elements (like street names or any other cognitive and social recognition of form elements, which depend on the social representation of urban space). A second additional aspect to be considered is the way distance between form elements is defined within the network: it can be topological (number of connections, whatever the nature of the connections is), angular (modelling the psychological impedance to change of direction in movement) or dimensional (metric or temporal distance on the network, modelling the physical impedance to movement). Table 1 is a systematic overview of different configurational techniques which are well known in literature (Cutini 2010), by positioning them in the three main methodological dimensions. A few combinations seem more problematic than others (as calculating an angular distance on a primal graph or a dimensional distance on a dual graph). Some techniques mix the options (even distance options could be mixed). More interestingly, primal approaches can easily integrate built-up elements in the configurational calculus (Sevtsuk et al. 2012). Although being mainly network-based, configurational analysis can thus integrate another important aspect of urban form.

Table 1. Overview Scheme of Configurational Approaches.

| GRAPH MODEL | | Primal | | | Dual | | | |
|---|---|---|---|---|---|---|---|---|
| | NETWORK DISTANCE | Topological | Angular | Dimensional | Topological | | Angular | Dimensional |
| ENTITIZING OF NETWORK ELEMENTS | Angular | | | | Axial Analysis (Hillier, Hanson, 1984) | (Porta, Latora, Cardillo, 2006) | Angular Analysis (Turner, 2000) | |
| | Topologic | Ma.P.P.A. * (Cutini, Petri, Santucci, 2004) | | M.C.A. * (Porta, Latora, Cardillo, 2006) | Road-centre Lines Analysis (Turner, 2007) | ICN | Continuity analysis (Figueiredo, Amorim, 2005) | |
| | Dimensional | | | M.C.A. * (Fusco, Caglioni, Araldi, 2016) | | | | |
| | Socially defined | | | | Street Name (Jiang, Claramunt, 2004) | | | |

\* Implementation with buildings possible    ▢ Techniques which will implemented in the paper

We will focus on three particularly well established techniques of configurational analysis: Space Syntax Axial Analysis (Axial SSx), Multiple Centrality Assessment (MCA) and Mark Point Parameter Analysis (MaPPA). The latter seems particularly interesting as it defines form elements by mixing topologic, angular and dimensional considerations: linear features are identified whenever an intersection occurs, a main directional change occurs (as in a winding street) or a given metric distance is achieved (which results in dividing a long street segment in several shorter ones). MCA entitization does not integrate changes of direction, another main difference being the network distance considered in the two techniques (topological vs. dimensional).

Beyond node degree (which has a quite different meaning in the primal and in the dual graph) every technique produces a series of comparable configurational indicators, directly derived from Freeman (1979) which can be calculated at different scales of analysis. Indicators are calculated for network points by the primal techniques of MCA and MaPPA and for axial lines by the dual technique of Axial SSx. The simplest indicator is the reach / node count, quantifying the number of form elements which can be reached from every element within a given radius of analysis. The farness / total depth indicator is a measure of how far a form element is from all other elements in a given radius. By normalizing it through the number of form elements within the same radius, we obtain the normalized farness / mean depth. The integration index of Axial SSx is just a further normalization of mean depth, with reference to values of reference theoretic street networks (Hillier 1996). The betweenness / choice indicator is a measure of how many shortest paths pass through a form element, when considering all shortest paths within a given radius. Reach or closeness (inverse farness) on the one side and betweenness on the other, represent two different and complementary aspects of centrality within a network: being near vs. being between the others.

Different urban morphologies should produce different distributions of configurational indicators among their form elements. Tree-like street networks should thus have very few elements with high betweenness / choice and many elements with much lower values. Grid plans or highly connective irregular plans of spontaneous urban morphologies should have more even distribution both of betweennes and farness indicators. We will thus test the selected configurational techniques in their ability to differentiate urban morphologies of different known characteristics.

## Results, Discussion and Conclusion

The city of Nice in southern France presents very diverse urban forms dues to its particular urban history (Graff 2013) where control of urban form and spontaneous growth have co-existed and shaped different city neighbourhoods. Six emblematic study areas within the city of Nice were thus selected (Figure 1) in order to represent specific patterns of urban form and functioning. This paper will focus on the cases of the city centre, mainly a planned urban grid of the second half of the XIX century, and of the western hills, a spontaneous growth area of the XX century marked by the presence of north/south valleys and ridges and a relatively marked tree-like network structure.

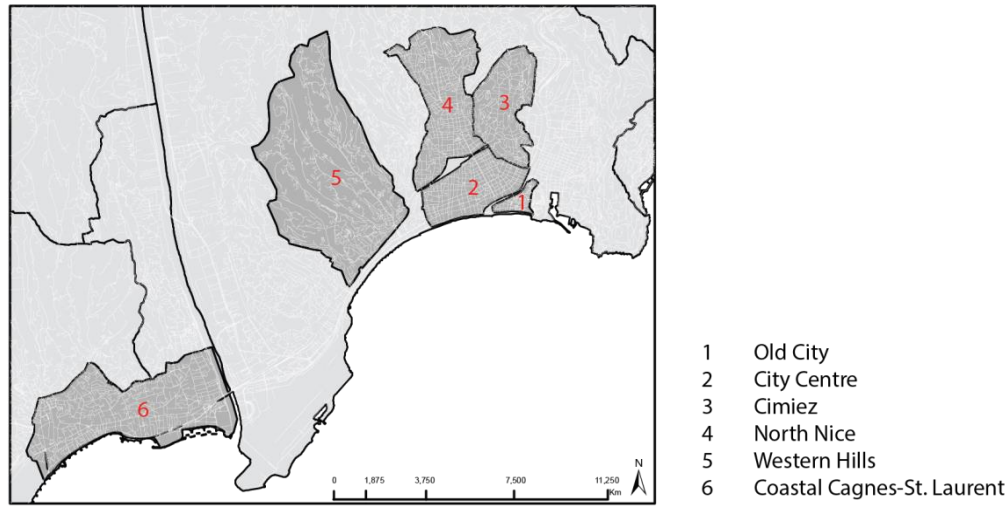

Figure 1. The city of Nice and its six emblematic study areas.

Five configurational calculi were carried out: Axial SSx, MCA and MaPPA, the latter two with and without buildings. MCA points are street intersections with additional points on segments exceeding 100 m. MaPPA points are street intersections, winding points and additional points on segments exceeding 100 m. When integrating buildings, MCA uses building points projected on the network, MaPPA weights its network points with the number of buildings assigned to them. Three scales of analysis were considered for each calculus: the micro-scale of 400 m/7 topological steps, the meso-scale of 800 m/15 steps, the macro-scale of 1600 m/30 steps. The configurational calculi were carried out on the whole built-up area of Nice and its neighbouring municipalities (including a wide buffer zone to avoid edge-effects). Results are nevertheless analysed locally for each emblematic areas (Figure 2): through maps with a relative colour scale (to identify the internal structure of each area) and through diagrams on an absolute scale defined by the range of the whole urban area of Nice (to compare the study areas among them).

Both the dual approach of Axial SSx and the primal approach of MaPPA can differentiate the two study areas, both internally and between them. They highlight however very different central features, above all in the city centre: the longest axial lines vs the many crossroads, which are much more evenly distributed in space. Axial SSx identifies highly skewed distributions (whether exponential or heavy-tailed, however with stronger hierarchy in the western hills area), whereas primal node degree distributions are more symmetrical. These results are direct consequences of the entitization difference of the dual approach. At the micro-scale, Axial SSx determines low values of reach in the western hills, when these are evaluated with respect to the whole urban area. Only the relative colour map can highlight the higher values of the southernmost parts, where the valleys meet. The city centre has a more symmetric distribution of values, like in the primal techniques, the longest lines constantly showing high values. MaPPA and MCA identify correctly the internal structure of the two areas both in the absolute and in the relative representation and are a bit more selective (especially MCA) in assigning high values within the city centre. The metric impedance of MCA also contributes to limit high centrality values in the southernmost sections of the western hills, whereas Axial SSx an MaPPA tend to stretch them further up the valleys. At the same time, MaPPA penalizes the most the hillsides, as they are connected with winding streets.

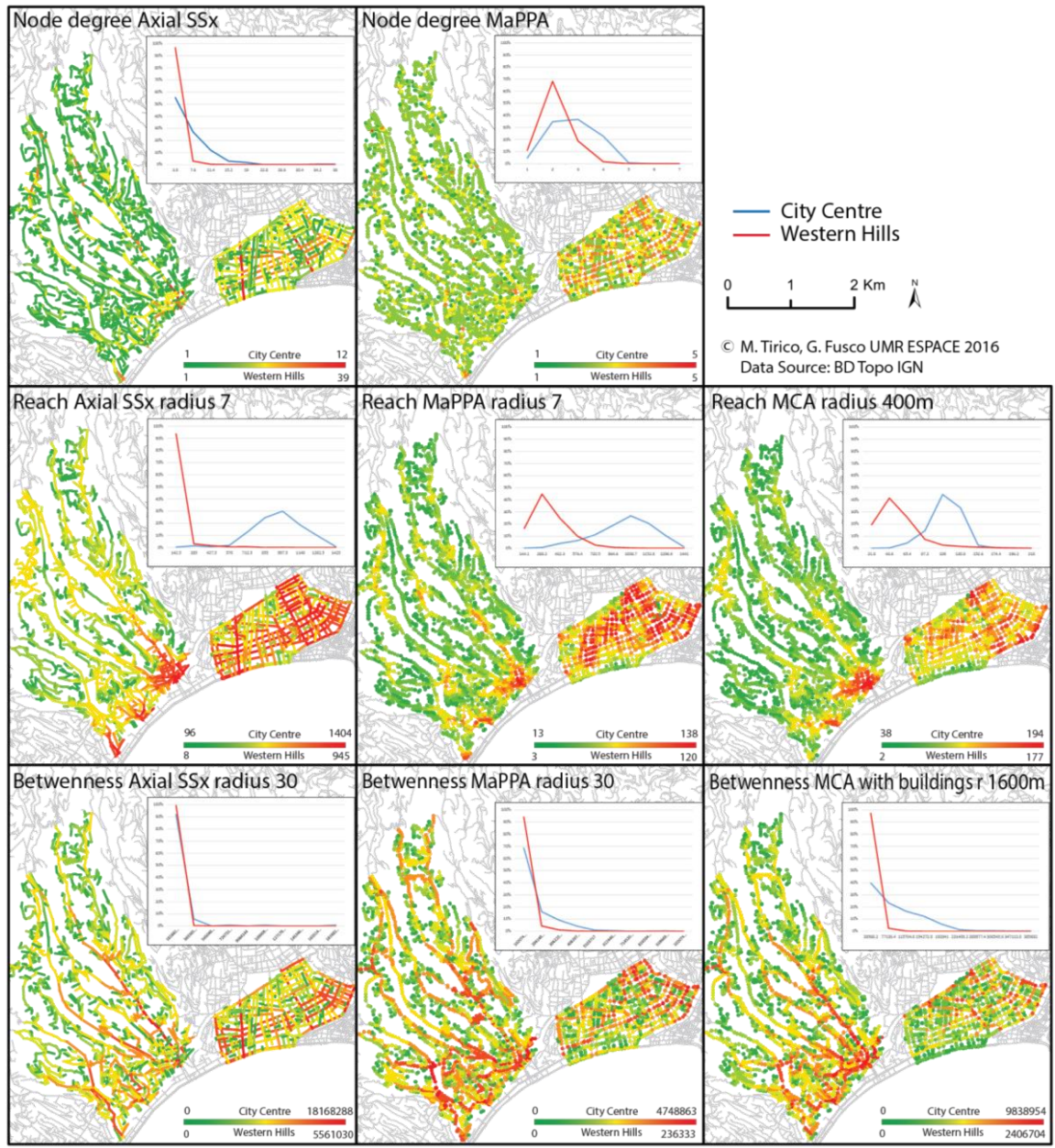

Figure 2. Empirical results for two emblematic areas in the city of Nice.

The large-scale betweenness of the Axial SSx fails to differentiate the city centre and the western hills, despite their very diverse network structures: both areas show extremely hierarchical distributions. Primal approaches of MaPPA and MCA (here implemented with buildings) more correctly show the less hierarchical distribution of values in the city centre. The inclusion of buildings contributes even more to the spreading of betweenness values, since the city centre, at the difference of the western hills, is homogenously built up. On the contrary, buildings highlight the betweenness centrality of thalweg streets, which are much more heavily urbanised than ridge streets.

In conclusion, as already pointed out by Porta et al. (2006a), results of Axial SSx seem more heavily dependent on the choices in entitizing the network. Axial SSx betweenness also fails to detect two particularly different urban morphologies in the city of Nice. Differences in the results of MaPPA and MCA, with and without the integration of buildings, are more subtle. The two techniques share the primal representation of the graph and the choice of adding dimensional constraints in the MaPPA entitization brings topological distance of MaPPA closer to dimensional distance of MCA. Introducing buildings seems to improve the detection of internal structure of case studies, as a further aspect of the morphology of urban fabric is, at least partially, injected in the configurational calculus of the network. Of course, a more systematic comparison of results is needed to confirm these conclusions, taking into consideration the six study areas and using more quantitative ways of comparing configurational results.